\documentclass{osa-article}

\journal{osac}

\articletype{Research Article}

\begin{document}

\title{Phase-dependent double optomechanically induced transparency in a hybrid optomechanical cavity system with coherently mechanical driving}

\author{Shi-Chao Wu,\authormark{1,3} Li-Guo Qin,\authormark{1,2,4} Jian Lu,\authormark{1} AND Zhong-Yang Wang\authormark{1,5}}

\address{\authormark{1}Shanghai Advanced Research Institute, Chinese Academy of Sciences, Shanghai 201210, China\\
\authormark{2}School of Mathematics, Physics and Statistics, Shanghai University Of Engineering Science, Shanghai 201620, China\\
\authormark{3}University of Chinese Academy of Sciences, Beijing 100049,China\\
\authormark{4}lgqin@foxmail.com\\
\authormark{5}wangzy@sari.ac.cn}


\begin{abstract}
  We propose a scheme that can generate a tunable double optomechanically induced transparency (OMIT) in a hybrid optomechanical cavity system, in which the mechanical resonator of an optomechanical cavity is coupled to an additional mechanical resonator, and the additional mechanical resonator can be driven by a weak external coherently mechanical driving field. We show that both the intensity and the phase of the external mechanical driving field can control the propagation of the probe field, including changing the transmission spectrum curve from a double-window to a single-window. Our study also provides an effective way to produce an intensity-controllable narrow-bandwidth transmission spectra, with the probe field modulated from excessive opacity to remarkable amplification.
\end{abstract}

\section{Introduction}

Engineering and manipulating the interaction between the optical and mechanical modes is an active research area, it has been studied theoretically and experimentally in many systems \cite{Natureultrafastconversion,andrewsNaturebidirectionalconver,NatureQuantumcoherent,naturenanomechanicalBochmann,NatCoherentBalram,RevMooptomechanicsAspelmeyer,Fan2015Cascaded,ScienceOptomechanicallyWeis,Xiao:14,Yang:17,XiongFundamentals}, in which a representative one of them is the optomechanical systems \cite{RevMooptomechanicsAspelmeyer,Fan2015Cascaded,ScienceOptomechanicallyWeis,Xiao:14,Yang:17,XiongFundamentals}. A traditional optomechanical system is composed of an optical cavity and a mechanical resonator. With the development of the micro- and nano-fabrication techniques, it is practicable to integrate the traditional optomechanical system with other systems, including the additional mechanical resonators \cite{Lin2010Coherent,PRATunableMa}, superconducting microwave cavities \cite{regalnaturemeasuring}, phononic \cite{Garcia2016Acoustic} or photonic crystal cavities \cite{naturenanomechanicalBochmann}, piezoelectric systems \cite{PRAppliedAcoustoOpticBalram} and charged systems \cite{PRAPrecisionZhang}. Compared with the traditional optomechanical system, the photon-phonon interaction in those hybrid optomechanical systems can be controlled by not only the optical radiation pressure, but also the piezoelectric forces \cite{naturenanomechanicalBochmann,NatCoherentBalram}, Lorentz forces \cite{Alexei2005Evidence,Xue2007Controllable} or Coulombic forces \cite{PRATunableMa,Wang2015Precision}.

The interaction between the optical and mechanical modes can generate many interesting phenomena in the optomechanical systems, such as the optomechanical induced transparency (OMIT). OMIT is a phenomenon that a photon-cavity can be changed from opacity to transparency, it arises from the quantum interference effect between different energy levels \cite{Zhang:18,PRAEITAgarwal,NatCoherentBalram,RMPHybridquantumXiang,ScienceOptomechanicallyWeis,safavinatureelectromagnetically,
PRAPhaseJia}. Similar to the electromagnetically induced transparency (EIT), which was observed in the three-level atomic systems \cite{PhysicsTodayHarris,PRLEITBoller,Xu:13,Cheng:17}, OMIT also can be been applied in many fields, including quantum ground-state cooling \cite{Bienert2015Optomechanical,Zhang:14}, fast and slow light \cite{2015tunablefast,Jiang:13}, and quantum information processing \cite{Stannigel2011Optomechanical,Mcgee2013Mechanical}.

Compared with a traditional optomechanical system, in the hybrid optomechanical systems, the single-OMIT has been extended to the double-OMIT \cite{Yang:17,Qin2015Tunable,PRAOptomechanicalWang,PRATunableMa,PRA2013PhononmediatedQu,PRAOptomechanicalWang,OC2014DoubleWen,Wu2018Microwave,PRAOptomechanicalWang}. Different from the single-window transmission spectrum observed in the OMIT, the double-window transmission spectrum can be observed in the double-OMIT. This phenomenon is similar to the two-photon absorption, which is observed in the four-level energy-structure atomic systems \cite{PRLPhotonHarris,OLNONABSORPMin}. In the double-OMIT hybrid optomechanical systems, which is combined of a traditional optomechanical systems and an additional two-level energy system, the original three-level energy system turned into a four-level energy system, under the the quantum interference between different energy level pathways, the double-OMIT is generated. Double-OMIT can be applied in many fields, including optical switches \cite{Wu2018Microwave}, temperature measurement \cite{Wang2015Precision}, high-resolution spectroscopy and double-channel quantum information processing \cite{PRATunableMa,PRAOptomechanicalWang}.

Here, we propose a hybrid optomechanical system, in which the mechanical resonator of an optomechanical cavity is coupled to an additional mechanical resonator, and the additional mechanical resonator can be driven by a weak external coherently mechanical driving field. In this system, a tunable double optomechanically induced transparency (OMIT) can be generate. We show that both the intensity and the phase of the external coherently mechanical driving field can control the propagation of the probe field, including changing the transmission spectrum curve from double-window to single-window.

This tunable double-OMIT phenomenon arises because that the energy-levels of the system can be modulated by the external mechanical driving field. In our system, a four-level energy structure can be formed, under the coupling effect between the two mechanical resonators, the original mechanical resonator energy level is dressed into an empty-level state. When the external driving field is applied to the additional mechanical resonator, the empty mechanical resonator level is replenished into an occupied-state. As a result, under the interference between the optical pumped field and the external mechanical driving field, the absorption and the dispersion of the probe field is modulated.

Experimentally and theoretically, many device can be applied to implement our scheme. With regard to the coupling of the two mechanical resonators, the two mechanical resonators can be coupled to each other via the common coupling overhang or the Coulomb interaction \cite {Qin2015Tunable,Barzanjeh2016Phonon,Xu2015Mechanical,Gil2011Exponential,Luo2018Strong,Karabalin2009Nonlinear,Pu2013Demonstration,Okamoto2013Coherent}. More importantly, the coupling strength between them can be modulated flexibly \cite{Pu2013Demonstration,Okamoto2013Coherent}. As to the mechanical driving field, there are many forms of driving forces can be applied, for example, the mechanical resonators can be driven by the piezoelectric forces, which a mechanical resonator is fabricated with piezoelectric materials \cite{PRACavityChangLZ,PRLMultimodeHan,PRLSuperconductingCleland,naturequantumConnell}; the mechanical resonators can also be driven by the Lorentz forces, which a current-carrying resonator is placed in the magnetic field \cite{Alexei2005Evidence,Xue2007Controllable}.

The hybrid optomechanical cavity systems with coherently mechanical driving has been studied in many fields, include realizing single \cite{Xu2015Controllable,Si2017Optomechanically,PRAPhaseJia} and double  \cite{Yang:17} optomechanically induced opacity and amplification, and weak-force measurement \cite{Ma2015Optomechanically}. Relative to those systems, our scheme has the following features: (i) our scheme can realize the switch between the double-OMIT and the single-OMIT, which can be controlled by many forms of the mechanical driving fields, such as the piezoelectric forces and Lorentz forces. (ii) Our study also provides an effective way to produce an intensity-controllable narrow-bandwidth transmission spectra, with the probe field modulated from excessive opacity to remarkable amplification.

This paper is organized as follows: In Sec. 2 we describe the theoretical model and derive the dynamical equation. In Sec. 3 we discuss the experimental feasibility and the physical mechanism of the double-OMIT. In Sec. 4 the external mechanical driving-field controlled double-OMIT is presented. The last section concludes the paper.

\section{Theoretical model and dynamical equation}

\begin{figure}[h!]
\centering\includegraphics[width=7cm]{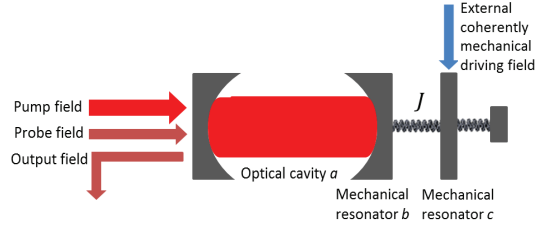}
\caption{(color online) The schematic diagram of the hybrid optomechanical system, in which the mechanical resonator $b$ of a traditional optomechanical cavity is coupled to an additional mechanical resonator $c$, and the additional mechanical resonator is driven by a weak external coherently mechanical driving field. The optomechanical cavity is driven by a strong optical pump field and a weak optical probe field. The coupling strength between the two mechanical resonators is referred to as $J$.}
\end{figure}

The schematic of the hybrid optomechanical cavity system we proposed is shown in Fig.1, in which the mechanical resonator $b$ of a traditional optomechanical cavity is coupled to an additional mechanical resonator $c$, and mechanical resonator $c$ is driven by a weak external coherently mechanical driving field. The frequency of the optical cavity is $\omega_a$, and the frequencies of the two mechanical resonators are $\omega_b$ and $\omega_c$, respectively. The mechanical resonators $b$ and $c$ can be coupled to each other via the common coupling overhang or Coulomb interaction \cite{Qin2015Tunable,Barzanjeh2016Phonon,Xu2015Mechanical,Gil2011Exponential,Luo2018Strong,Karabalin2009Nonlinear,Pu2013Demonstration,Okamoto2013Coherent}. The mechanical driving field can be applied to the system in the forms of piezoelectric forces \cite{PRACavityChangLZ,PRLMultimodeHan,PRLSuperconductingCleland,naturequantumConnell} or Lorentz forces \cite{Alexei2005Evidence,Xue2007Controllable}.
As a result, the hybrid optomechanical cavity system can be driven by both the optical fields and the mechanical driving field. We assume that the optomechanical cavity is driven by a strong optical pump field with frequency $\omega_{pu}$ and a weak optical probe field with frequency $\omega_{pr}$. Mechanical resonator $c$ is driven by a weak external coherently mechanical driving field with frequency $\omega_{d}$. The optomechanical coupling strength between the optical fields and the mechanical resonator $b$ is referred to as $G_{om}$, and the coupling strength between the mechanical resonators $b$ and $c$ is referred to as $J$. In the frame rotating at the frequency of the pump field $\omega_{pu}$, the Hamiltonian of the total system is given by:
\begin{eqnarray}
H=H_{0}+H_{I},
\end{eqnarray}
where
\begin{eqnarray}
H_0=\hbar{\Delta}_{{a}}{\hat{a}}^{\dag}a+\hbar\omega_{b}{\hat{b}}^{\dag}\hat{b}+\hbar{\omega}_{c}{{\hat{c}}}^{\dag}\hat{c},
\end{eqnarray}
\begin{eqnarray}
H_I=-\hbar{g_{om}}\hat{a}^{\dag}\hat{a}(\hat{b}^{\dag}+\hat{b})-\hbar{J}(\hat{b}^{\dag}\hat{c}+\hat{b}\hat{c}^{\dag})
\nonumber\\+({i}{\hbar}{\varepsilon}_{pu}\hat{a}^{\dag}+{i}{\hbar}{\varepsilon}_ {pr}e^{-i{\delta}t-i{\phi_p}}{\hat{a}}^{\dag}
 +{i}{\hbar}{\varepsilon}_{d}e^{-i{\omega_d}t-i{\phi_d}}\hat{c}^{\dag}+H.c.).
\end{eqnarray}
$H_0$ is the free Hamiltonian of the system, where $\hat{a}$, $\hat{b}$ and $\hat{c}$ are the annihilation operators of the optical cavity mode $a$, the mechanical phonon modes $b$ and $c$, respectively. $\Delta_a=\omega_a-\omega_{pu}$ is the frequency-detuning of the optical pump field from the optical cavity. $H_I$ describes the interaction Hamiltonian. The first term is the optomechanical interaction term, $g_{om}=({\omega_a}/{L})\sqrt{\hbar/m\omega_b}$ denotes the single-photon optomechanical coupling strength between the optical field mode and the mechanical phonon mode, where $m$ is the effective mass of mechanical mode and $L$ is the effective length of the optical cavity. The second term is the interaction term between the coupling mechanical resonators $b$ and $c$, where $J$ is the coupling strength. The last four terms describe the energy of the input fields, $\varepsilon_{pu}={\sqrt{P_{pu}\kappa_{a}/(\hbar\omega_{pu})}}$ and $\varepsilon_{pr}={\sqrt{P_{pr}\kappa_{a}/(\hbar\omega_{pr})}}$ are the amplitudes of the optical pump field and probe field, respectively, where $P_{pu}$ and $P_{pr}$ are the input powers of the optical pump field and optical probe field, respectively. $\kappa_a$ is the decay rates of the optical cavity. $\delta=\omega_{pr}-\omega_{pu}$ is the frequency-detuning of the optical probe field from the pump field, and ${\phi_p}$ is the phase-difference between the probe field and pump field. The amplitude, frequency and the phase of the external mechanical driving field are defined as $\varepsilon_{d}$, $\omega_{d}$, and ${\phi_d}$.

By adopting the quantum Langevin equations (QLEs) for the operators, in which the damping and noise terms are supplemented \cite{PRAPhaseJia,PRAEntanglingBarzanjeh}, we can get:
\begin{eqnarray}
\begin{split}
\dot{\hat{a}}=-(i\Delta_{a}+\frac{\kappa_a}{2})\hat{a}+ig_{om}{\hat{a}}({\hat{b}}^{\dag}+{\hat{b}})
+\varepsilon_{pu}+\varepsilon_{pr}{e^{-i{\delta t}-i{\phi_p}}}+\hat{f}_{in},\\
\dot{{\hat{b}}}=-(i\omega_{b}+\frac{\gamma_{b}}{2}){\hat{b}}+ig_{om}{\hat{a}^\dag}\hat{a}+iJ{\hat{c}}+\hat{\varsigma}_{in},\\
\dot{\hat{c}}=-(i\omega_{c}+\frac{\gamma_c}{2}){\hat{c}}+iJ{\hat{b}}+\varepsilon_{d}e^{-i{\omega_d}t-i{\phi_d}}+\hat{\xi}_{in},
\end{split}
\end{eqnarray}
where $\gamma_{b}$ and $\gamma_{c}$ are the intrinsic damping rates of mechanical resonator $b$ and $c$, respectively; $\hat{f}_{in}$ is the optical input noise; $\hat{\varsigma}_{in}$ and $\hat{\xi}_{in}$ are the quantum Brownian noises acting on the mechanical resonators $b$ and $c$, respectively \cite{PRAEntanglingBarzanjeh}. For simplicity, the hat symbols of the operators are omitted in the following description.

Relative to the intensity of the optical pump field, we assume that both the intensities of the optical probe field and the weak external mechanical driving field satisfies the conditions $\varepsilon_{pr} \ll\varepsilon_{pu}$ and $\varepsilon_{d} \ll\varepsilon_{pu}$. Then we linearize the dynamical equations of the system by assuming $a={a}_{s}+\delta{a}$, $b={b}_{s}+\delta{b}$ and $c={c}_{s}+\delta{c}$, all of which are composed of an average amplitude and a fluctuation term. ${a}_{s}$, ${c}_{s}$ and ${b}_{s}$ are steady-state values when only the strong optical pump field is applied. Assuming $\varepsilon_{pr} \rightarrow0$, $\varepsilon_{d} \rightarrow0$ and setting all the time derivatives to zero, we can get
\begin{eqnarray}
\begin{split}
{a}_{s}=\frac{\varepsilon_{pu}}{i\Delta_{a}^{\prime}+\frac{\kappa_a}{2}},\\
{b}_{s}=\frac{ig_{om}{|a_s|}^2+iJ{c}_{s}}{i\omega_{b}+\frac{\gamma_{b}}{2}},\\
{c}_{s}=\frac{iJ{b}_{s}}{i\omega_{c}+\frac{\gamma _c}{2}},
\end{split}
\end{eqnarray}
where $\Delta_{a}^{\prime}=\Delta_{a}-g_{om}( {b}_{s}^{\ast}+{b}_{s} )$, $\Delta_{a}^{\prime}$ is the effective frequency-detuning of the optical pump field from the optical cavity, including the frequency shift caused by the mechanical motion. Furthermore, by substituting the assumptions $a={a}_{s}+\delta{a}$, $b={b}_{s}+\delta{b}$ and $c={c}_{s}+\delta{c}$ into the nonlinear QLEs and dropping the small nonlinear terms, we can obtain the linearized QLEs as follows:
\begin{eqnarray}
\begin{split}
\dot{\delta{{a}}}=-(i\Delta_a^{\prime}+\frac{\kappa_a}{2})\delta{a}+iG_{om}(\delta{b^{\dag}}+\delta{b})+{\varepsilon}_{pr}e^{-i{\delta}t-i{\phi_p}}+f_{in},\\
\dot{\delta{{b}}}=-(i\omega_{b}+\frac{\gamma_{b}}{2})\delta{b}+i(G_{om}^{\ast}{\delta{a}}+G_{om}\delta{a^{\dag}})+iJ\delta{c}+\varsigma_{in},\\
\dot{\delta{{c}}}=-(i\omega_{c}+\frac{\gamma_c}{2})\delta{c}+iJ\delta{b}+{\varepsilon}_ {d}e^{-i{\omega_d}t-i{\phi_d}}+\xi_{in},
\end{split}
\end{eqnarray}
where $G_{om}=g_{om}{{a_{s}}}$ is the total coupling strength between the optical cavity mode $a$ and mechanical mode $b$.

 We assume that the cavity is driven by the optical pump field at the red sideband $\Delta_a^{\prime}=\omega_{b}$, and the system is operated in the resolved sideband regime, in which $\omega_b(\omega_{c})\gg\kappa_a$. The mechanical resonator also has a high quality factor for $\omega_b(\omega_{c})\gg\gamma_b (\gamma_{c})$, the mechanical frequency $\omega_{b}$ $(\omega_{c})$ is also much larger than $G_{om}$. For simplify, we also assume that the frequencies of the mechanical resonators $b$ and $c$ are equal, and the frequency of the weak external mechanical driving field is also equal with mechanical $c$, which $\omega_d=\omega_b=\omega_c$. With $\delta=\omega_d$, then the fluctuation terms $\delta{a}$, $\delta{b}$, $\delta{c}$ and the noise terms ${f_{in}}$, ${\varsigma_{in}}$, ${\xi_{in}}$ can be rewritten as
\begin{eqnarray}
\begin{split}
\delta{a}=\delta{a_{+}}e^{-i{\delta}t}+\delta{a_{-}}e^{i{\delta}t},\\
\delta{b}=\delta{b_{+}}e^{-i{\delta}t}+\delta{b_{-}}e^{i{\delta}t},\\
\delta{c}=\delta{c_{+}}e^{-i{\delta}t}+\delta{c_{-}}e^{i{\delta}t},\\
\delta{f_{in}}=\delta{{f_{in}}_{+}}e^{-i{\delta}t}+\delta{{f_{in}}_{-}}e^{i{\delta}t},\\
\delta{\varsigma_{in}}=\delta{{\varsigma_{in}}_{+}}e^{-i{\delta}t}+\delta{{\varsigma_{in}}_{-}}e^{i{\delta}t},\\
\delta{\xi_{in}}=\delta{{\xi_{in}}_{+}}e^{-i{\delta}t}+\delta{{\xi_{in}}_{-}}e^{i{\delta}t},
\end{split}
\end{eqnarray}
where $\delta{O_{+}}$ and $\delta{O_{-}}$ (with $O=a,b,c$) correspond to the components at the original frequencies of $\omega_{pr}$ and $2\omega_{pu}-\omega_{pr}$, respectively \cite{PRAEITSumei,PRAPrecisionZhang}. Next we substituted Eq. (7) into Eq. (6) and ignored the second-order small terms, by equating coefficients of terms with the same frequency, the component at the frequencies $\omega_{pr}$ can be obtained as:
\begin{eqnarray}
\begin{split}
\dot{\delta{{a_{+}}}}=(i\lambda_a-\frac{\kappa_a}{2})\delta{a_{+}}+iG_{om}\delta{b_{+}}+{\varepsilon}_{pr}e^{-i{\phi_p}}+{f_{in}}_{+},\\
\dot{\delta{{b_{+}}}}=(i\lambda_{b}-\frac{\gamma_{b}}{2})\delta{b_{+}}+iG_{om}^{\ast}{\delta{a}}_{+}+iJ\delta{c}_{+}+{\varsigma_{in}}_{+},\\
\dot{\delta{{c_{+}}}}=(i\lambda_c-\frac{\gamma_c}{2})\delta{c_{+}}+iJ\delta{b}_{+}+{\varepsilon}_{d}e^{-i{\phi_d}}+{\xi_{in}}_{+},
\end{split}
\end{eqnarray}
where $\lambda_a=\delta-\Delta_{a}^{\prime}$, $\lambda_{b}=\delta-\omega_{b}$, and $\lambda_c=\delta-\omega_{c}$, they satisfy the relation $\lambda_a=\lambda_{b}=\lambda_c$.

For the noise terms, they obey the following correlation fluctuations\cite{RevMooptomechanicsAspelmeyer}:
\begin{eqnarray}
\begin{split}
\langle{{{{\hat{f}}_{in}(t) \hat{f}}_{in}^{\dag}(t^{\prime})}}\rangle=[N(\omega_{a})+1]\delta(t-t^{\prime}),\\
\langle{{{{\hat{f}}_{in}^{\dag}(t) \hat{f}}_{in}(t^{\prime})}}\rangle=[N(\omega_{a})]\delta(t-t^{\prime}),\\
\langle{{{{\hat{\varsigma}}_{in}(t)\hat{\varsigma}}_{in}^{\dag}(t^{\prime})}}\rangle=[N(\omega_{b})+1]\delta(t-t^{\prime}),\\
\langle{{{{\hat{\varsigma}}_{in}^{\dag}(t)\hat{\varsigma}}_{in}(t^{\prime})}}\rangle=[N(\omega_{b})]\delta(t-t^{\prime}),\\
\langle{{{{\hat{\xi}}_{in}(t) \hat{\xi}}_{in}^{\dag}(t^{\prime})}}\rangle=[N(\omega_{c})+1]\delta(t-t^{\prime}),\\
\langle{{{{\hat{\xi}}_{in}^{\dag}(t) \hat{\xi}}_{in}(t^{\prime})}}\rangle=[N(\omega_{c})]\delta(t-t^{\prime}),
\end{split}
\end{eqnarray}
where $N(\omega_{a})=[exp(\hbar\omega_{a}/k_{B}T)-1]^{-1}$ is the equilibrium mean thermal photon numbers of the optical fields; $N(\omega_{b})=[exp(\hbar\omega_{b}/k_{B}T)-1]^{-1}$ and $N(\omega_{c})=[exp(\hbar\omega_{c}/k_{B}T)-1]^{-1}$ are the equilibrium mean thermal phonon numbers of the mechanical resonators $b$ and $c$. To neglect the influences of noise, we assume that our system is operated in the low-enough temperature, which simultaneously satisfying the conditions $\hbar\omega_{a}/k_{B}T\gg1$, $\hbar\omega_{b}/k_{B}T\gg1$ and $\hbar\omega_{c}/k_{B}T\gg1$, correspondingly, $\langle{{{f_{in}}_{+}}}\rangle=\langle{{\varsigma_{in}}_{+}}\rangle=\langle{{\xi_{in}}_{+}}\rangle=0$ \cite{RevMooptomechanicsAspelmeyer,Photonsfri}. Under the mean-field steady-state condition $\langle\dot{\delta{{a_{+}}}}\rangle=\langle\dot{\delta{{b_{+}}}}\rangle=\langle\dot{\delta{{c_{+}}}}\rangle=0$, we can obtain
\begin{eqnarray}
\begin{split}
0=(i\lambda_a-\frac{\kappa_a}{2})\langle\delta{a_{+}}\rangle+iG_{om}\langle\delta{b_{+}}\rangle+\varepsilon_{pr}e^{-i{\phi_p}},\\
0=(i\lambda_{b}-\frac{\gamma_{b}}{2})\langle\delta{b_{+}}\rangle+iG_{om}^{\ast}\langle{\delta{a_{+}}}\rangle+iJ\langle\delta{c_{+}}\rangle,\\
0=(i\lambda_c-\frac{\gamma_c}{2})\langle\delta{c_{+}}\rangle+iJ\langle\delta{b_{+}}\rangle+\varepsilon_{d}e^{-i{\phi_d}}.
\end{split}
\end{eqnarray}

The solution of $\langle\delta{a_{+}}\rangle$ can be obtained as
\begin{flushleft}
\begin{eqnarray}
\begin{split}
\langle\delta{a_{+}}\rangle
=e^{-i{\phi_p}}[\frac{-G_{om}J\varepsilon_{d}e^{-i{\phi}}}{(\frac{\kappa_a}{2}-i\lambda_a)[{(\frac{\gamma_b}{2}-i\lambda_b)({\frac{\gamma_{c}}{2}
-i\lambda_c})+{J^2}}]+{{|{G_{om}}|}^2}({\frac{\gamma_{c}}{2}
-i\lambda_c})}\\
+\frac{[{(\frac{\gamma_b}{2}-i\lambda_b)({\frac{\gamma_{c}}{2}
-i\lambda_c})+{J^2}}]\varepsilon_{pr}}{(\frac{\kappa_a}{2}-i\lambda_a)[{(\frac{\gamma_b}{2}-i\lambda_b)({\frac{\gamma_{c}}{2}
-i\lambda_c})+{J^2}}]+{{|{G_{om}}|}^2}({\frac{\gamma_{c}}{2}
-i\lambda_c})}].
\end{split}
\end{eqnarray}
\end{flushleft}
where $\phi=\phi_d-\phi_p$ is the phase-difference between the external mechanical driving field and the optical fields. Based on the input-output relation, the output field at the probe frequency $\omega_{pr}$ can be expressed as \cite{PRAEITAgarwal,PRAPrecisionZhang}
\begin{eqnarray}
\varepsilon_{out}=\kappa_{a} \langle\delta{a_{+}}\rangle-\varepsilon_{pr}e^{-i{\phi_p}}.
\end{eqnarray}
The transmission coefficient $T_{pr}$ of the probe field is given by \cite{ScienceOptomechanicallyWeis,PRAPhaseJia}
\begin{eqnarray}
T_{pr}=\frac{\varepsilon_{out}}{\varepsilon_{pr}e^{-i{\phi_p}}}=\kappa_{a} \langle\delta{a_{+}}\rangle/{\varepsilon_{pr}e^{-i{\phi_p}}}-1.
\end{eqnarray}
Defining $\varepsilon_{T}=\frac{{\kappa_a}{\langle\delta{a_{+}}\rangle}}{\varepsilon_{pr}e^{-i{\phi_p}}}$, we obtain the quadrature $\varepsilon_{T}$ of the output field at the frequency $\omega_{pr}$:

\begin{eqnarray}
\begin{split}
\varepsilon_{T}={\kappa_{a}}[\frac{-G_{om}J{\eta}e^{-i{\phi}}}{(\frac{\kappa_a}{2}-i\lambda_a)[{(\frac{\gamma_b}{2}-i\lambda_b)({\frac{\gamma_{c}}{2}
-i\lambda_c})+{J^2}}]+{{|{G_{om}}|}^2}({\frac{\gamma_{c}}{2}
-i\lambda_c})}\\
+\frac{[{(\frac{\gamma_b}{2}-i\lambda_b)({\frac{\gamma_{c}}{2}
-i\lambda_c})+{J^2}}]}{(\frac{\kappa_a}{2}-i\lambda_a)[{(\frac{\gamma_b}{2}-i\lambda_b)({\frac{\gamma_{c}}{2}
-i\lambda_c})+{J^2}}]+{{|{G_{om}}|}^2}({\frac{\gamma_{c}}{2}
-i\lambda_c})}],
\end{split}
\end{eqnarray}
where $\eta={\varepsilon_{d}}/{\varepsilon_{p}}$ is the amplitude-ratio between the external mechanical driving field and the optical probe field. The real part Re $[\varepsilon_T]$ and imaginary part Im $[\varepsilon_T]$ describe the absorption and dispersion of the system, respectively.

If we assume that there is no external mechanical driving field applied to the additional mechanical resonator $c$, defining $\lambda=\lambda_a=\lambda_b$, after the simplification, the term $\varepsilon_{T}$ becomes
\begin{eqnarray}
\varepsilon_{T}=\frac{\kappa_{a}}{\frac{\kappa_a}{2}-i{\lambda}+\frac{A_{+}}{\lambda_{+}-i{\lambda}}+\frac{A_{-}}{\lambda_{-}-i{\lambda}}},
\end{eqnarray}
where $\lambda_{\pm}$ and $A_{\pm}$ are
\begin{eqnarray}
\begin{split}
\lambda_{\pm}=\frac{\frac{\gamma_{b}}{2}+\frac{\gamma_{c}}{2}{\pm}i{\sqrt{{4{J}^2-(\frac{\gamma_b}{2}-\frac{\gamma_{c}}{2})^2}}}}{2},\\
A_{\pm}=\pm\frac{\lambda_{\pm}-\frac{\gamma_{c }}{2}}{\lambda_{+}-\lambda_{-}}{{|G_{om}|}}^2.
\end{split}
\end{eqnarray}
This expression has the standard form for the double-OMIT, which is similar to the double-EIT \cite{PRADoubleAlotaibi}. If we eliminate the coupling interaction between the mechanical resonators $b$ and $c$, $\varepsilon_T$ becomes
\begin{eqnarray}
\varepsilon_{T}=\frac{\kappa_{a}}{\frac{\kappa_a}{2}-i\lambda_a+\frac{|{G_{om}}|^2}{\frac{\gamma_b}{2}-i\lambda_b}},
\end{eqnarray}
it has the standard form of the single-OMIT window, which is similar to the standard EIT \cite{PhysicsTodayHarris}.

\section{Physical mechanism of the system}

\begin{figure}[h!]
\centering\includegraphics[width=7cm]{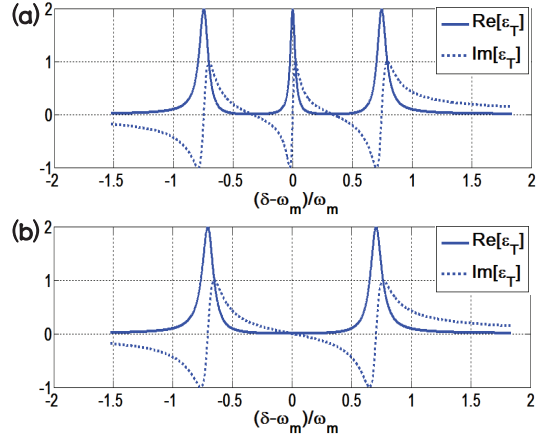}
\caption{ (a)The absorption Re$[\varepsilon_T]$ (solid line) and dispersion Im$[\varepsilon_T]$ (dashed line)as function of $(\delta-\omega_{m})/\omega_{m}$ for the double-OMIT, under the condition that there is no external mechanical driving field applied. With the parameters $g_{om}/{2\pi}=2.7$ Hz, $J/{2\pi}=320$ KHz, $\kappa_{a}/{2\pi}=215$ KHz, $\omega_{a}/{2\pi}=194$ THz, $\omega_{m}/{2\pi}=\omega_{b}/{2\pi}=\omega_{c}/{2\pi}=947$ KHz, $\kappa_{b}/{2\pi}=\kappa_{c}/{2\pi}=140$ Hz, $P_{pu}=1$ mW, and $\Delta_a=\omega_b=\omega_c$. (b) The absorption Re$[\varepsilon_T]$ (solid line) and dispersion Im$[\varepsilon_T]$ (dashed line) as function of $(\delta-\omega_{m})/\omega_{m}$ for the single-OMIT, under the condition $J=0$, the other parameters are the same as those in Fig.2 (a).}
\end{figure}

\begin{figure}[htbp]
\centering\includegraphics[width=7cm]{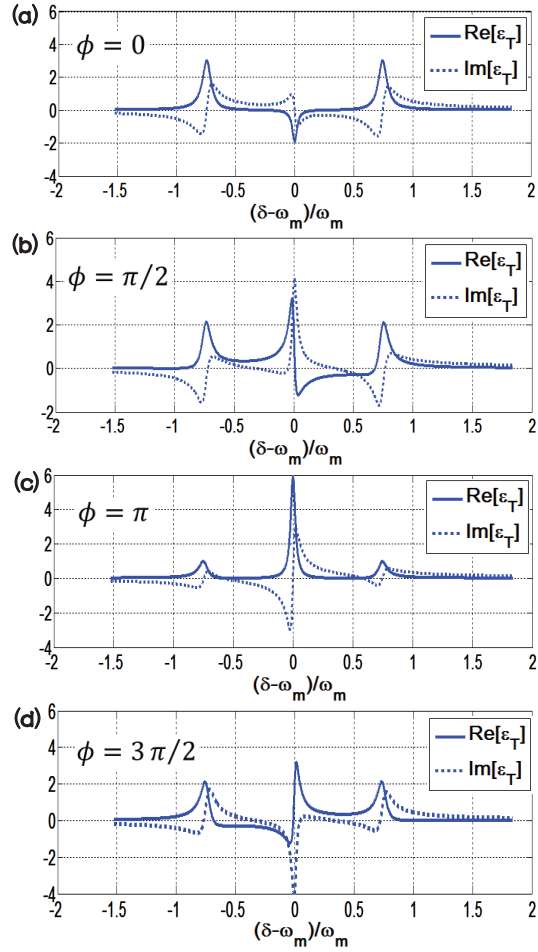}
\caption{ The absorption Re$[\varepsilon_T]$ (solid line) and dispersion Im$[\varepsilon_T]$ (dashed line)as function of $(\delta-\omega_{m})/\omega_{m}$ for different phase factors: (a) $\phi=0$; (a) $\phi=\pi/2$; (a) $\phi=\pi$; (a) $\phi={3\pi}/2$. Here $\eta=1$, the other parameters are the same as those in Fig.2 (a).}
\end{figure}

\begin{figure}[htbp]
\centering\includegraphics[width=7cm]{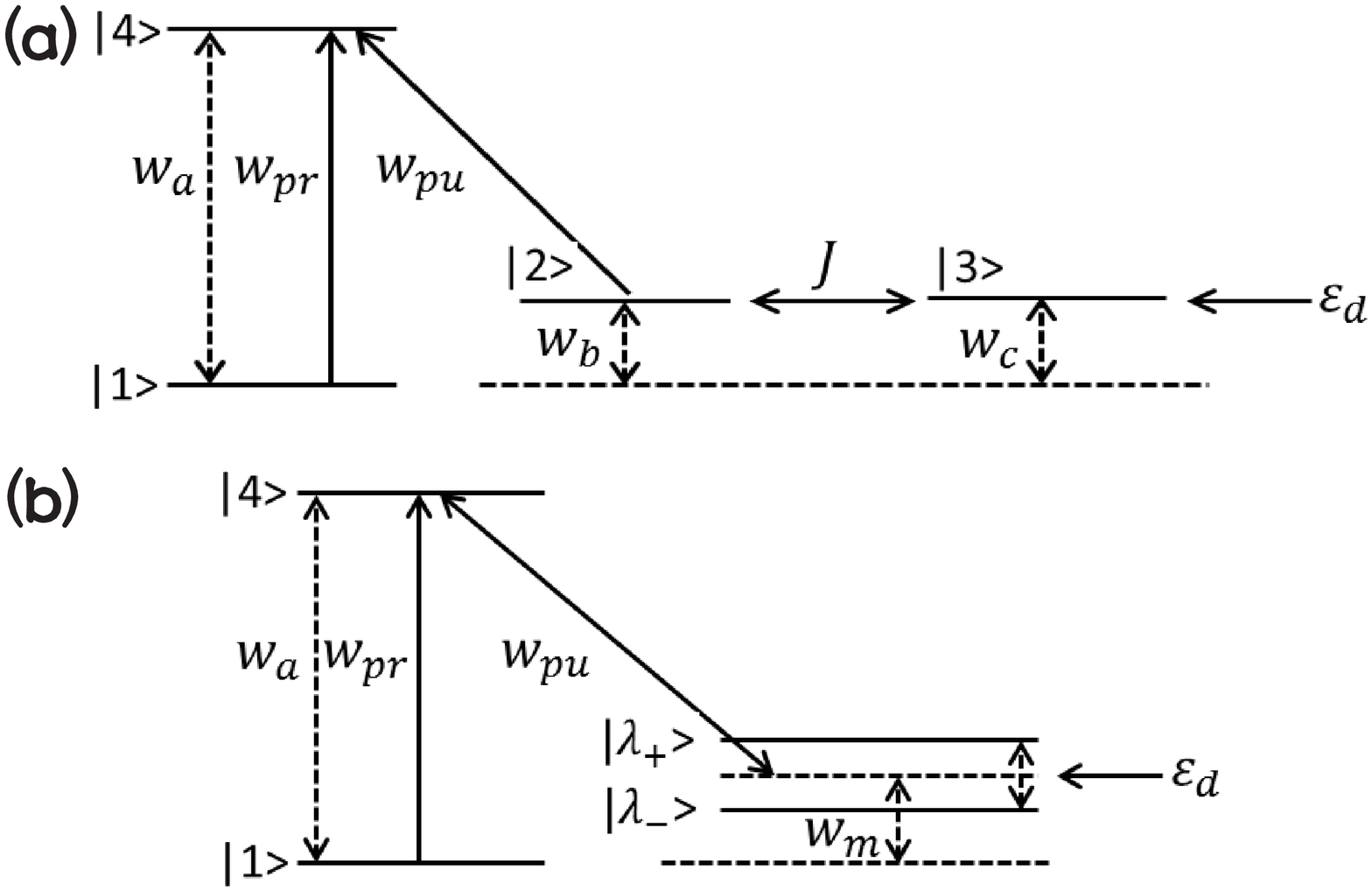}
\caption{(a) Energy level structure of the hybrid optomechanical system, where the states of the levels are designated as $|i\rangle (i=1, 2, 3, 4)$. The amplitude of the external mechanical driving field is $\varepsilon_{d}$. The energy differences between $|{1}\rangle$ and $|{4}\rangle$, $|{1}\rangle$ and $|{2}\rangle$, and $|{1}\rangle$ and $|{3}\rangle$ are the frequencies of cavity $a$, mechanical resonator $b$, and mechanical resonator $c$, which are designated as $\omega_a$, $\omega_b$ and $\omega_c$, respectively. $\omega_{pr}$ is equal to $\omega_{a}$, and the detuning events between them are $\omega_{pr}-\omega_{pu}=\omega_{b}=\omega_{c}$. (b) Energy level structure of the hybrid optomechanical system in the dressed-state picture. With the relation $\omega_{m}=\omega_{b}=\omega_{c}$. $|\lambda_{\pm}\rangle$ are the new dressed levels generated by the coupling effect between the two mechanical resonators, with the energy difference of $2J$.}
\end{figure}

We present below a discussion of the feasibility of the tunable double-OMIT in the hybrid optomechanical cavity system. In our system, the parameters we chosen are shown as below, which are all based on a realistic system \cite{Simon2009Observation}. The frequency and the decay rate of the optical cavity are $\omega_{a}/{2\pi}=194$ THz and $\kappa_{a}/{2\pi}=215$ KHz. For simplify, we assume the frequencies of the mechanical resonators $b$ and $c$ are equal,which $\omega_{m}/{2\pi}=\omega_{b}/{2\pi}=\omega_{c}/{2\pi}=947$ KHz. They both have high quality factors for $\gamma_b/{2\pi}=\gamma_c/{2\pi}=140$ Hz, which strictly meet the condition of the sideband-resolved regime $\omega_{b}\gg \kappa_{a}$. The single-photon optomechanical coupling strength is $g_{om}/2\pi = 2.7$ Hz. Furthermore, we assume the coupling strength between the mechanical resonator $b$ and $c$ is $J/{2\pi}=320$ KHz \cite{Lin2010Coherent}. Next, we consider the situation that the optical cavity is driven by the optical pump field at the mechanical red sideband, where $\Delta_{a}^{\prime}=\omega_{b}$.

When there is no external driving applied to the mechanical resonators $c$, as shown in Fig. 2(a), the absorption Re $[\varepsilon_T]$ and dispersion Im $[\varepsilon_T]$ of the optical probe field as function of $(\delta-\omega_{m})/\omega_{m}$ are plotted. In the transmission spectrum curve, a double-transparency window can be obtained, and the positions of two minima points are determined by the imaginary part of $\lambda_{\pm}$, as shown in Eq.(16). The distance between the two minima points is $2J$, which is closely related to the coupling strength between the mechanical resonators $b$ and $c$. If there is no coupling interaction between the mechanical resonators $b$ and $c$, as shown in Fig. 2(b), a single-transparency window in the transmission spectrum curve can be obtained. In this situation, the minima point is determined by the frequency point $\delta=\omega_m$, and the relevant mechanisms have also been studied extensively \cite{PRAEITAgarwal}.

Now we consider the situation that a weak external coherently mechanical driving field is applied to the system. For simplify, We assume that the frequency of the mechanical driving field is equal with mechanical $c$, which $\omega_{d}=\omega_{c}$, then we have $\omega_{d}=\omega_{m}=\omega_{b}=\omega_{c}$, and we also assumed that the amplitude-ratio between the external mechanical driving field and the optical probe field meet the condition $\eta={\varepsilon_{d}}/{\varepsilon_{p}}=1$. As shown in Figs. 3(a)-3(d), we plot the absorption Re $[\varepsilon_T]$ and dispersion Im $[\varepsilon_T]$ of the optical probe field as function of $(\delta-\omega_{m})/\omega_{m}$ for different phase-differences $\phi$. When $\phi=0$ and $\phi=\pi$, the absorption and the dispersion curves of them are symmetric and antisymmetry, respectively. The difference between the situations $\phi=0$ and $\phi=\pi$ is that when $\phi=0$, the destructive interference between the two terms in Eq. (14) suppressed the absorption at the frequency position $\delta=\omega_m$; when $\phi=\pi$, the constructive interference between the two terms in Eq. (14) amplified the absorption at the frequency position $\delta=\omega_m$. When $\phi=\pi/2$ and $\phi={3\pi}/2$, the absorption and dispersion curves of them are both anomalous, and the maximum points of the absorption and the dispersion curves appear in the red- or blue-detuned regions, respectively. Especially, comparing the situations when $\phi=\pi/2$ and $\phi={3\pi}/2$, it is shown that the absorption curves between them are mirror-symmetry.

As to the standard double-transparency window, which is shown in Figs. 2(a), it is originates from the quantum interference effect between different energy level pathways. In the hybrid system, a four-level energy configuration is formed by the energy levels of the optical cavity and the mechanical resonators. Under the coupling effect between the mechanical resonators $b$ and $c$, the original mechanical resonator energy $\omega_{m}$ level is split into two new dressed levels. With $\Delta_{a}^{\prime}=\omega_{m}=\omega_b=\omega_{c}$, the two new dressed levels are $\lambda_{\pm}$, and the disparity between them is $2J$, as shown in the dressed-state picture in Fig. 4 (b). Under the effects of the optical radiation pressure, the quantum interference between different energy level pathways occurs, the third-order nonlinear absorption is enhanced by the constructive quantum interference and the linear absorption is inhibited by the destructive quantum interference. As a result, the double-OMIT window appears, and the relevant mechanisms have been studied extensively \cite{PRAObservationYan,PRLCrossPhaseLi}.

Compared Fig. 3 with Fig. 2, it is shown that when a weak external mechanical driving field applied to the mechanical resonator $c$, the absorption and the dispersion at the frequency position $\delta=\omega_m$ is modulated obviously by the phase of the external mechanical driving field. This phenomenon arises because the energy levels of the system is modulated by the external mechanical driving field. As shown in Figs. 4(b), under the coupling between the mechanical resonators $b$ and $c$, the original mechanical resonator $\omega_{m}$ level turned into an empty-level state. When the external mechanical driving field applied to the mechanical resonator $c$, with the condition that the frequency of the mechanical driving field is equal to the the frequency of mechanical resonator, the empty mechanical resonator level is replenished into an occupied-state by the mechanical driving field. As a result, under the interference between the optical pumped field and the mechanical driving field, the absorption and the dispersion near the frequency position $\delta=\omega_m$ is modulated, which is correspondingly consistent with the Eq. (14).

\section{Tunable double-OMIT of the system}

\begin{figure}[htbp]
\centering\includegraphics[width=7cm]{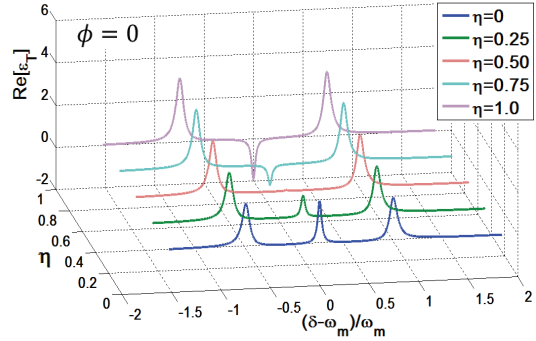}
\caption{(color online) The absorption Re$[\varepsilon_T]$ as functions of $(\delta-\omega_{b})/\omega_{b}$ and $\eta$ (units of $\eta$ is $0.25$), which under the condition $\phi=0$, the other parameters are the same as in Fig. 2 (a).}
\end{figure}

\begin{figure}[htbp]
\centering\includegraphics[width=7cm]{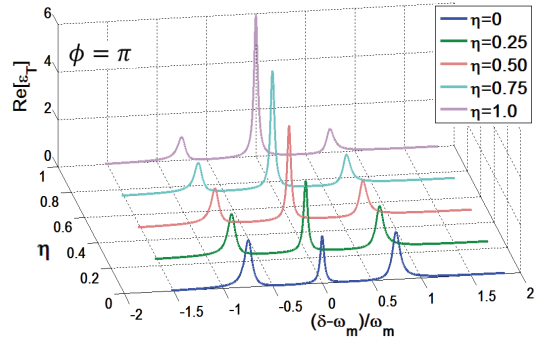}
\caption{(color online) The absorption Re$[\varepsilon_T]$ as functions of $(\delta-\omega_{b})/\omega_{b}$ and $\eta$ (units of $\eta$ is $0.25$), which under the condition $\phi=\pi$, the other parameters are the same as in Fig. 2 (a).}
\end{figure}

To further explore the effect of the external mechanical driving field, we plot the absorption Re$[\varepsilon_T]$ as functions of $(\delta-\omega_{m})/\omega_m$ and $\eta$, which $\eta={\varepsilon_{d}}/{\varepsilon_{p}}$ is the amplitude-ratio between the mechanical driving field and the optical probe field. We consider the situation in which the phase-difference between the mechanical driving field and the optical fields meet the condition $\phi=0$, as shown in Fig. 5. In the absence of the mechanical driving field, which $\eta=0$, under the interference between the optical pump field and the optical probe field, a standard double-OMIT window appears. With the enhancement of the mechanical driving intensity, the absorption-rate at the frequency position $\delta=\omega_m$ is decreased from positive to negative gradually, which the narrow probe curve at the frequency position $\delta=\omega_m$ is modulated from full-opacity to remarkable amplification. More importantly, when $\eta=0.5$, the  transmission curve converts to a standard single-OMIT window, which the probe spectra at the frequency position $\delta=\omega_m$ is transmitted perfectly. This phenomenon arises from the destructive quantum interference between the optical fields and the mechanical driving field, which is consistent with the Eq. (14).

Furthermore, we discuss the situation in which the phase-difference between the mechanical driving field and the optical fields meet the condition $\phi=\pi$. As shown in Fig. 6, which is the absorption Re$[\varepsilon_T]$ as functions of $(\delta-\omega_{b})/\omega_b$ and $\eta$. In the absence of the mechanical driving field, the transmission spectrum curve is a standard double-OMIT window. With the enhancement of the mechanical driving intensity, the absorption-rate at the frequency position $\delta=\omega_m$ is amplified, which the narrow probe curve at the frequency position $\delta=\omega_m$ is modulated from full-opacity to excessive opacity, as a result, the intensity of the probe field in the cavity is enhanced. This phenomenon is opposite to the situation showed in Fig. 5, it originates from the constructive quantum interference between the optical fields and the mechanical driving field, which is also consistent with the Eq. (14).

Based on the Fig. 5 and Fig. 6, it shows that when the phase of the external mechanical driving field meet the conditions $\phi=0$ or $\phi=\pi$, the absorption-rate at the frequency position $\delta=\omega_m$ is proportionable with the intensity of the external driving field. As a result, the narrow probe curve can be modulated from excessive opacity to remarkable amplification, this phenomenon can be applied in many fields. For example, as the transmission spectrum curve can be changed from the double-window to the single-window, our system can be applied to realize the double-channel quantum information processing and optical switch, and the similar phenomena have been studied in the natural atomic systems \cite{PRLPairedphotonsBali,PRAswitchingKumar}. Moreover, the line-width of the narrow transmission curve at frequency $\delta=\omega_{m}$ is approximately equal to $J$, which is small enough relative to the line-width of the optical cavity, as a result, our system can be potentially applied in the field of tunable high-resolution spectroscopy, which is similar to the sub-Doppler spectral resolution observed in natural atomic systems \cite{PRLNonlinearHarris}.

\section{Conclusion}

In conclusion, the hybrid optomechanical cavity system we proposed provides a feasible way to control the double-OMIT by a weak external mechanical driving field. In this system, the empty-state dressed by the coupling effect between the mechanical resonators can be replenished into an occupied-state by the external coherently mechanical driving field. Under the interference between the optical pumped field and the external mechanical driving field, the absorption and the dispersion of the probe field is modulated. It is shown that both the intensity and the phase of the external coherently mechanical driving field can control the propagation of a probe field, including changing the transmission spectrum from double-window to single-window. Our system can be used to realize the tunable high-resolution spectroscopy and optical switch, more importantly, our system can be extended to other hybrid solid-state systems for exploring new quantum phenomena.

\section*{Funding}
Strategic Priority Research Program (XDB01010200); Hundred Talents Program of the Chinese Academy of Sciences (Y321311401); National Natural Sciences Foundation of China (61605225, 11674337 and 11547035); Natural Science Foundation of Shanghai (16ZR1448400).


\bibliography{doubleOMIT}





\end{document}